\documentclass[sigconf,screen]{acmart}
\renewcommand\footnotetextcopyrightpermission[1]{} 
\usepackage{subcaption}
\usepackage{multirow}
\usepackage{array}
\usepackage{booktabs}
\usepackage{tabularx}
\usepackage{makecell}
\usepackage{fancyhdr}

\AtBeginDocument{%
  }

\newcounter{MingyuanNumberOfComments}
\stepcounter{MingyuanNumberOfComments}

\setcopyright{acmlicensed}

\copyrightyear{2026} 
\acmYear{2026} 
\setcopyright{rightsretained}
\acmConference[MM '26]{Proceedings of the 34th ACM International Conference
on Multimedia}{November 10--14, 2026}{Rio de Janeiro, Brazil}
\acmBooktitle{Proceedings of the 34th ACM International Conference on
Multimedia (MM '26), November 10--14, 2026, Rio de Janeiro,
Brazil}

\acmSubmissionID{128}

\begin{document}

\title{QoS-QoE Translation with Large Language Model}



\author{Yingjie Yu}
\email{yyu69@illinois.edu}
\affiliation{%
  \institution{University of Illinois Urbana-Champaign}
  \city{Urbana}
  \state{Illinois}
  \country{USA}
}

\author{Mingyuan Wu}
\email{mw34@illinois.edu}
\affiliation{%
  \institution{University of Illinois Urbana-Champaign}
  \city{Urbana}
  \state{Illinois}
  \country{USA}
}

\author{Ahmadreza Eslaminia}
\email{ae15@illinois.edu}
\affiliation{%
  \institution{University of Illinois Urbana-Champaign}
  \city{Urbana}
  \state{Illinois}
  \country{USA}
}

\author{Lingzhi Zhao}
\email{lz26@illinois.edu}
\affiliation{%
  \institution{University of Illinois Urbana-Champaign}
  \city{Urbana}
  \state{Illinois}
  \country{USA}
}

\author{Kaizhuo Yan}
\email{kaizhuo2@illinois.edu}
\affiliation{%
  \institution{University of Illinois Urbana-Champaign}
  \city{Urbana}
  \state{Illinois}
  \country{USA}
}

\author{Klara Nahrstedt}
\email{klara@illinois.edu}
\affiliation{%
  \institution{University of Illinois Urbana-Champaign}
  \city{Urbana}
  \state{Illinois}
  \country{USA}
}
\renewcommand{\shortauthors}{Yingjie Yu et al.}
\begin{abstract}

QoS-QoE translation is a fundamental problem in multimedia systems because it characterizes how measurable system and network conditions affect user-perceived experience. Although many prior studies have examined this relationship, their findings are often developed for specific setups and remain scattered across papers, experimental settings, and reporting formats, limiting systematic reuse, cross-scenario generalization, and large-scale analysis. To address this gap, we first introduce \textit{QoS-QoE Translation} dataset, a source-grounded dataset of structured QoS-QoE relationships from the multimedia literature, with a focus on video streaming related tasks. We construct the dataset through an automated pipeline that combines paper curation, QoS-QoE relationship extraction, and iterative data evaluation. Each record preserves the extracted relationship together with parameter definitions, supporting evidence, and contextual metadata. We further evaluate the capability of large language models (LLMs) on QoS-QoE translation, both before and after supervised fine-tuning on our dataset, and show strong performance on both continuous-value and discrete-label prediction in bidirectional translation, from QoS-QoE and QoE-QoS. Our dataset provides a foundation for benchmarking LLMs in QoS-QoE translation and for supporting future LLM-based reasoning for multimedia quality prediction and optimization. The complete dataset and code are publicly available at \url{https://yyu6969.github.io/qos-qoe-translation-page/}, for full reproducibility and open access.

\end{abstract}



\maketitle
\thispagestyle{plain}
\fancyhead{}
\renewcommand{\headrulewidth}{0pt}
\pagestyle{plain}
\section{Introduction}

\begin{figure*}
    \centering
    \includegraphics[width=\textwidth]{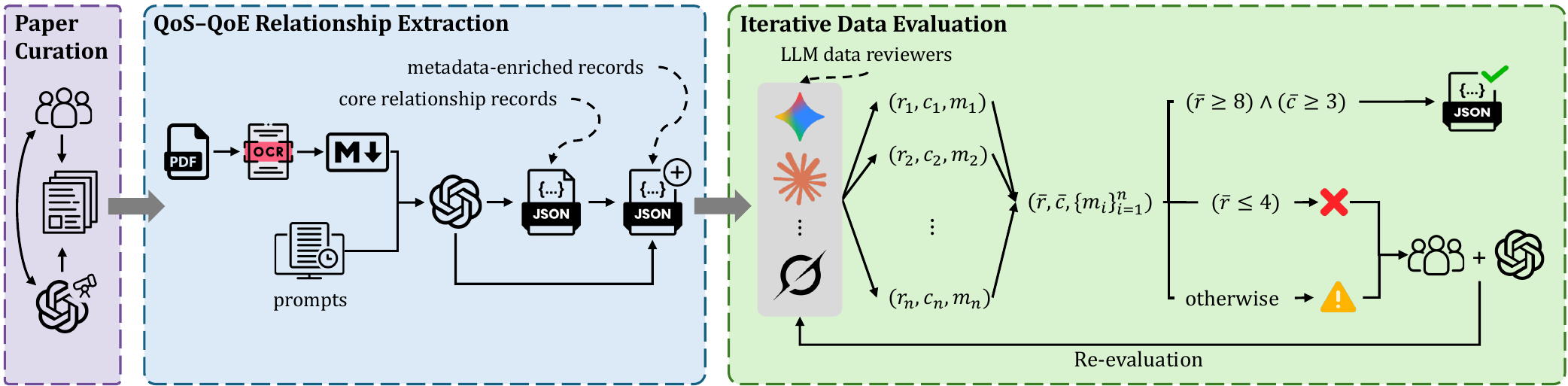}
    \caption{Overview of the QoS-QoE Translation dataset construction pipeline. The pipeline begins with paper curation, followed by QoS-QoE relationship extraction and iterative data evaluation.}
    \label{fig:data_pipeline}
\end{figure*}

Quality of Service (QoS) and Quality of Experience (QoE) are two central concepts in multimedia systems. QoS describes measurable system, network, and service conditions such as bitrate, delay, jitter, packet loss, and rebuffering, while QoE reflects users' perceived quality of the delivered service \cite{itu_gstr_rq_2023}. Understanding the QoS-QoE relationship is important for multimedia applications because it supports system design, adaptive streaming, network optimization, and user-centered quality prediction \cite{alreshoodi2013survey,barakabitze2020qoe}.

A large body of prior work has studied QoS-QoE relationships in multimedia applications, especially video streaming, by modeling how QoS factors map to perceived QoE, and in some cases how QoE targets guide adaptation decisions, using methods such as subjective experiments, heuristic rules, analytical modeling, and machine learning-based prediction \cite{alreshoodi2013survey,barman2019qoe,mao2017neural,yin2015control}. These studies have clarified how factors such as bitrate adaptation, stalling, startup delay, resolution changes, and network impairments affect perceived quality. However, many of these approaches are developed for particular setups and validated under specific conditions, which makes their findings and models difficult to generalize across scenarios. Applying them to new settings often requires substantial re-modeling, additional measurements, or new subjective studies.

These limitations motivate a more unified QoS-QoE translation capability that can support both forward translation from QoS to QoE and reverse translation from QoE targets to QoS conditions across diverse scenarios. Achieving this goal requires both strong models and high-quality data. LLM-based systems are a promising foundation because they have shown strong potential in multimedia-related tasks such as video understanding, audio processing, and multimodal agent-style decision making, while also supporting flexible reasoning and structured prediction \cite{huang2024survey_eval_mllm,li2024mvbench,tang2023salmonn,liu2024visualagentbench}. At the same time, constructing suitable source-grounded data is challenging because QoS-QoE relationships are scattered across the literature, reported in heterogeneous forms such as text, tables, figures, and equations, and often accompanied by incomplete or implicit contextual metadata.

To address this gap, we present \textit{QoS-QoE Translation} dataset, a source-grounded dataset of structured QoS-QoE relationships from the literature, with a current focus on video streaming. Our goal is to transform prior studies into a reusable data resource for the multimedia community. Each entry preserves the extracted relationship together with supporting evidence and contextual metadata, enabling interpretability and reproducibility. Because reported QoS-QoE relationships rarely appear in a single uniform format and often must be recovered from multiple forms of source evidence together with their surrounding context, we construct the dataset through a pipeline that combines paper curation, QoS-QoE relationship extraction, and iterative data evaluation, as shown in Figure~\ref{fig:data_pipeline}. This design supports large-scale dataset construction while maintaining quality control and traceability. To assess the utility of \textit{QoS-QoE Translation}, we perform supervised fine-tuning (SFT) of large language models (LLMs) on bidirectional QoS-QoE translation tasks and evaluate both continuous value and discrete label prediction. Results show strong performance gain, with the best fine-tuned model achieving 90.24\% Accuracy for discrete label prediction and 8.49\% MAPE (Mean Absolute Percentage Error) for continuous value prediction. These findings suggest that \textit{QoS-QoE Translation} provides a strong foundation for training LLMs to reason about QoS-QoE relationships and opens up new opportunities for applying LLM in multimedia applications.

The main contributions of this work are three-fold: 1) We introduce \textit{QoS-QoE Translation}, a source-grounded dataset of structured QoS-QoE relationships from the literature, with a current focus on video streaming. 2) We develop a reusable dataset construction pipeline for paper curation, relationship extraction, metadata enrichment, and iterative multi-reviewer quality evaluation. 3) We demonstrate that the dataset supports effective SFT of LLMs for bidirectional QoS-QoE translation and are the first to benchmark existing open-source LLMs in this domain.
\section{Dataset Construction}

Figure~\ref{fig:data_pipeline} overviews the \textit{QoS-QoE Translation} construction pipeline, which includes paper curation, QoS-QoE relationship extraction, and iterative data evaluation. Together, these stages transform curated papers into structured records and improve their quality through iterative review. Although \textit{QoS-QoE Translation} focuses on video streaming, the pipeline is reusable and can be adapted to other application domains that require extracting source-grounded relationships from the literature.

\subsection{Paper Curation}

We begin by constructing a curated corpus of research papers on QoS-QoE relationships in video streaming. To identify relevant and high-quality studies, we combine human screening with OpenAI deep research-assisted literature search \cite{openai_deep_research_2025}. We focus on papers published between 2017 and 2025, since advances in streaming systems, codecs, devices, and network configurations can change the practical meaning of reported QoS-QoE relationships over time. This restriction emphasizes recent and practically relevant evidence while reducing noise from older system settings. Using this human-AI curation process, we collect 505 papers related to QoS-QoE relationships in video streaming, which serve as the foundation for the downstream extraction pipeline and improve the relevance and reliability of the final dataset.

\subsection{QoS-QoE Relationship Extraction}
Starting from the curated paper corpus, we perform QoS-QoE relationship extraction from academic papers. Because the source papers are provided in PDF format, we first use MinerU \cite{wang2024mineru}, an OCR-based document parsing tool, to convert them into machine-readable markdown while preserving textual content and document structure for downstream processing. The converted content, together with carefully designed prompts, is then provided to an LLM for structured information extraction.

We use GPT-5.2 Thinking \cite{openai_gpt52_2025} as the core extraction model because its reasoning ability and long-context support make it well suited for source-grounded extraction from complex academic papers. During development, it provided a practical balance between extraction quality and cost for large-scale dataset construction.

This extraction stage produces two levels of outputs. The first is a set of core relationship records, which capture the fundamental QoS-QoE relationships extracted from source evidence such as equations, tables, and figures. The second is a set of metadata-enriched records, which augment the core relationship records with contextual metadata such as protocol, network type, device type, and scenario. By separating core relationship extraction from contextual metadata enrichment, the pipeline keeps the extracted relationships grounded in source evidence while still providing richer context for downstream analysis and reuse.

\subsection{Iterative Data Evaluation}

To improve data quality, we further introduce an iterative data evaluation stage. As shown in Figure~\ref{fig:data_pipeline}, each metadata-enriched record is reviewed by multiple LLM-based data evaluators. In the current dataset construction setup, we instantiate this stage with three data reviewers: Gemini-2.5-flash-lite~\cite{google2025gemini25flashlite}, Claude-haiku-4-5-20251001~\cite{anthropic2025claudehaiku45}, and Grok-4.20-0309-reasoning~\cite{xai_grok4200309reasoning}. Each reviewer provides a rating, a confidence score, and written feedback, denoted in the figure as a tuple of the form $(r_i, c_i, m_i)$. The rating score $r_i$ is assigned on a 0--10 scale, where 0, 2, 4, 6, 8, and 10 denote strong reject, reject, weak reject, weak accept, accept, and strong accept, respectively. The confidence score $c_i$ is assigned on a 1--5 scale, where higher values indicate stronger reviewer confidence in the judgment. Reviewer comments are also required to describe the identified issues and suggest possible solutions.

These reviewer outputs are aggregated into a decision using the average rating $\bar{r}$, the average confidence $\bar{c}$, and the collection of reviewer feedback messages. Based on our empirical inspection of reviewer outputs during dataset construction, we found the following thresholds to provide a reasonable balance between retaining high-quality records and filtering out unreliable extractions.
\begin{equation}
\text{Decision} =
\begin{cases}
\textbf{Accept}, & \text{if } \bar{r} \geq 8 \text{ and } \bar{c} \geq 3, \\
\textbf{Reject}, & \text{if } \bar{r} \leq 4, \\
\textbf{Conditional Accept}, & \text{otherwise.}
\end{cases}
\end{equation}
Records that satisfy the accept condition are retained as valid JSON entries, while records that satisfy the reject or conditional accept conditions are sent to a re-evaluation stage. In this stage, human guidance and an LLM are jointly used to revise the data before returning it to the evaluation loop. This iterative mechanism reduces unsupported, ambiguous, or low-quality extraction results and improves the consistency of the final dataset.
\section{Dataset Overview and Analysis}

\begin{figure}
    \centering
    \includegraphics[width=\columnwidth]{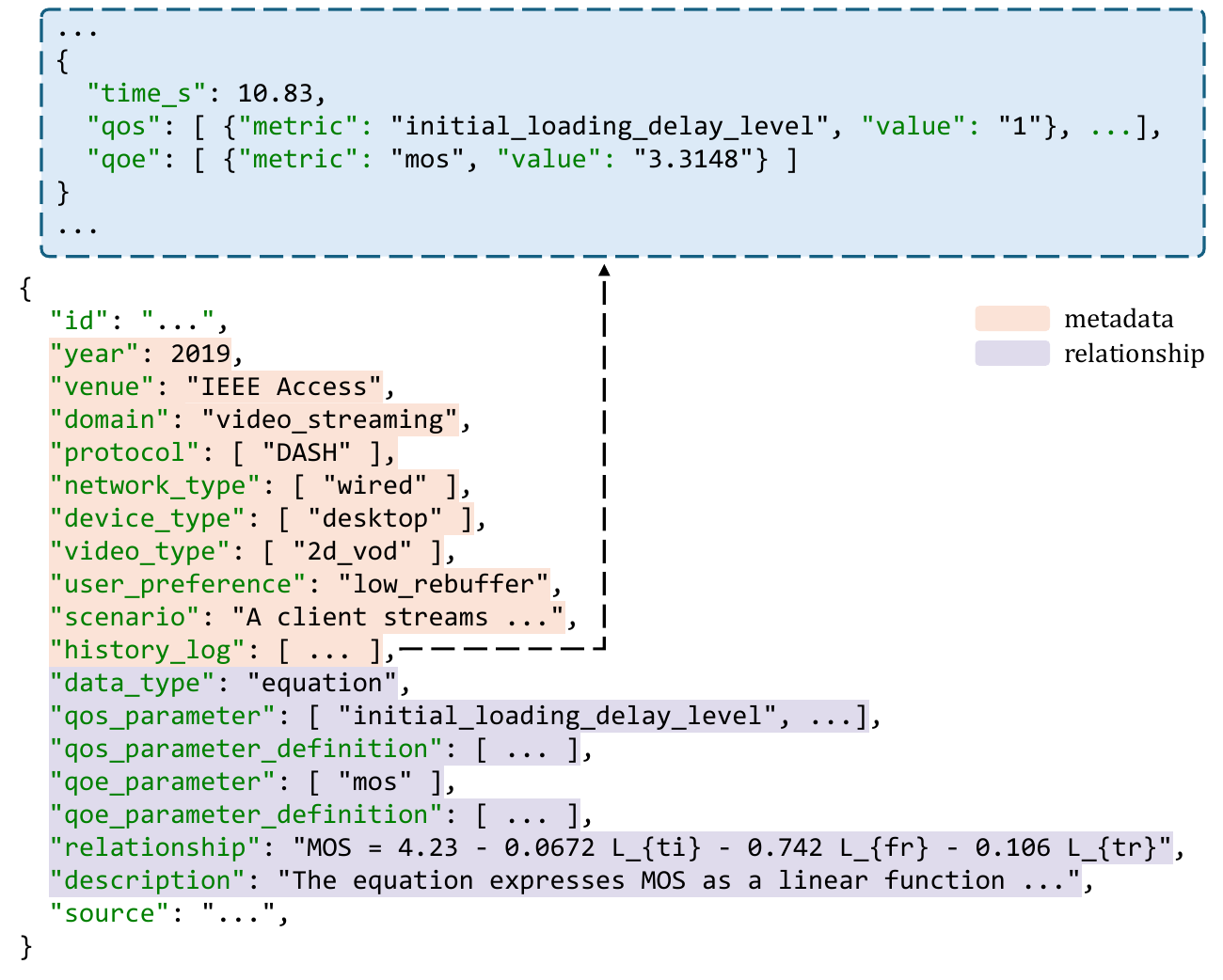}
    \caption{An example JSON record from the QoS-QoE Translation, showing enriched metadata and core relationship.}
    \label{fig:data_example}
\end{figure}

\begin{table}[t]
\centering
\caption{Field definitions of QoS-QoE Translation.}
\label{tab:dataset_field_definitions}
\footnotesize
\setlength{\tabcolsep}{4pt}
\renewcommand{\arraystretch}{1.1}
\begin{tabular}{>{\raggedright\arraybackslash}p{0.34\columnwidth} >{\raggedright\arraybackslash}p{0.60\columnwidth}}
\toprule
\textbf{Field} & \textbf{Definition} \\
\midrule
\texttt{id} & Unique identifier for each dataset record. \\
\texttt{year} & Publication year of the source paper. \\
\texttt{venue} & Publication venue of the source paper. \\
\texttt{domain} & Application domain of the record. \\
\texttt{protocol} & Streaming or transport protocol used in the study. \\
\texttt{network\_type} & Access network environment. \\
\texttt{device\_type} & Client device used for content consumption. \\
\texttt{video\_type} & Video content category or media format. \\
\texttt{user\_preference} & User preference emphasized in the study. \\
\texttt{scenario} & Summary of the experimental or evaluation setting. \\
\texttt{history\_log} & Temporally ordered QoS-QoE observations. \\
\texttt{data\_type} & Evidence type, such as equation, table, or figure. \\
\texttt{qos\_parameter} & QoS variables in the relationship. \\
\texttt{qos\_parameter\_definition} & Definitions of the QoS variables. \\
\texttt{qoe\_parameter} & QoE variables in the relationship. \\
\texttt{qoe\_parameter\_definition} & Definitions of the QoE variables. \\
\texttt{relationship} & Extracted dependency between QoS and QoE. \\
\texttt{description} & Natural-language explanation of the relationship. \\
\texttt{source} & Source-grounded evidence trace. \\
\bottomrule
\end{tabular}
\end{table}

\begin{figure*}[t]
    \centering

    \begin{subfigure}[t]{0.24\textwidth}
        \centering
        \includegraphics[width=\textwidth]{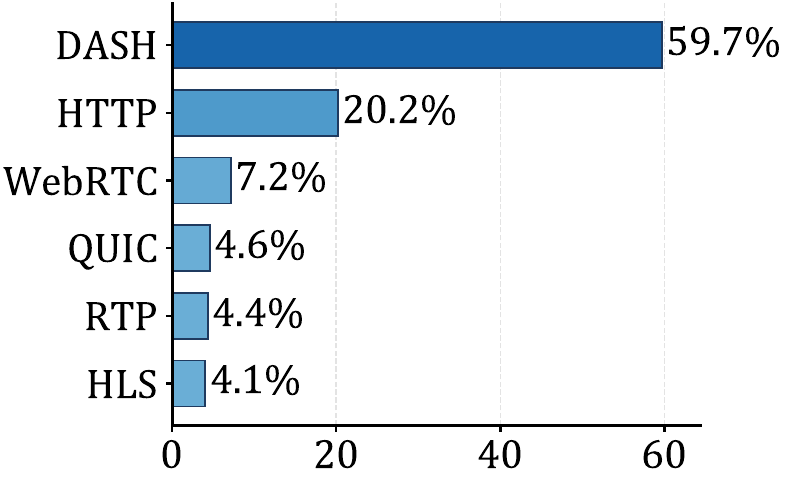}
        \caption{Protocol}
        \label{fig:protocol_horizontal_bar}
    \end{subfigure}
    \hfill
    \begin{subfigure}[t]{0.24\textwidth}
        \centering
        \includegraphics[width=\textwidth]{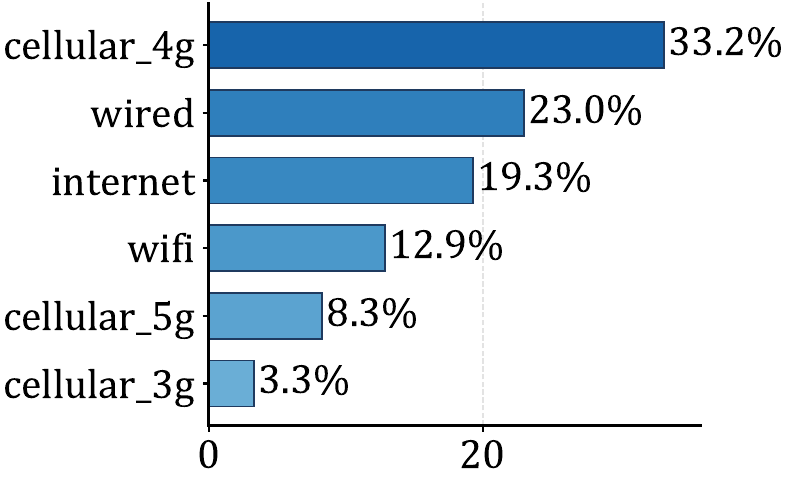}
        \caption{Network type}
        \label{fig:network_type_horizontal_bar}
    \end{subfigure}
    \hfill
    \begin{subfigure}[t]{0.24\textwidth}
        \centering
        \includegraphics[width=\textwidth]{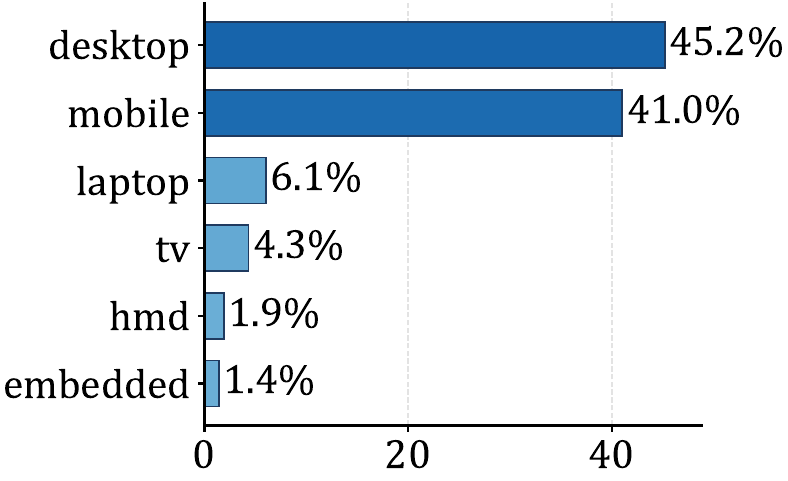}
        \caption{Device type}
        \label{fig:device_type_horizontal_bar}
    \end{subfigure}
    \hfill
    \begin{subfigure}[t]{0.24\textwidth}
        \centering
        \includegraphics[width=\textwidth]{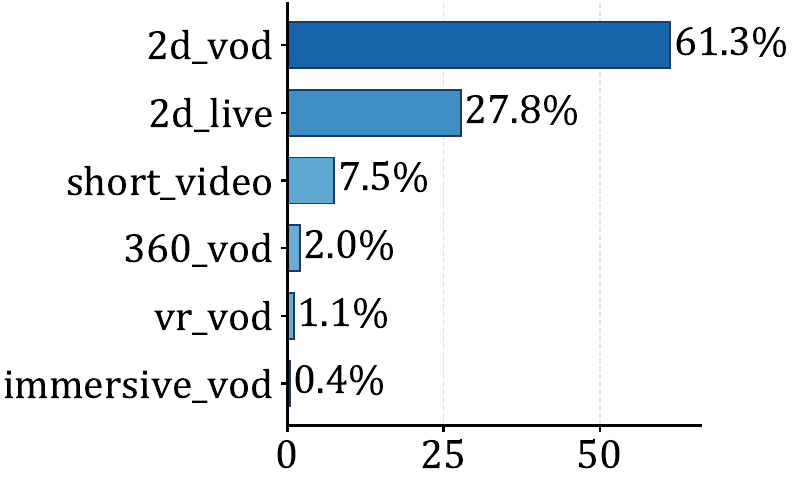}
        \caption{Video type}
        \label{fig:video_type_horizontal_bar}
    \end{subfigure}

    \vspace{0.8em}

    \begin{subfigure}[t]{0.24\textwidth}
        \centering
        \includegraphics[width=\textwidth]{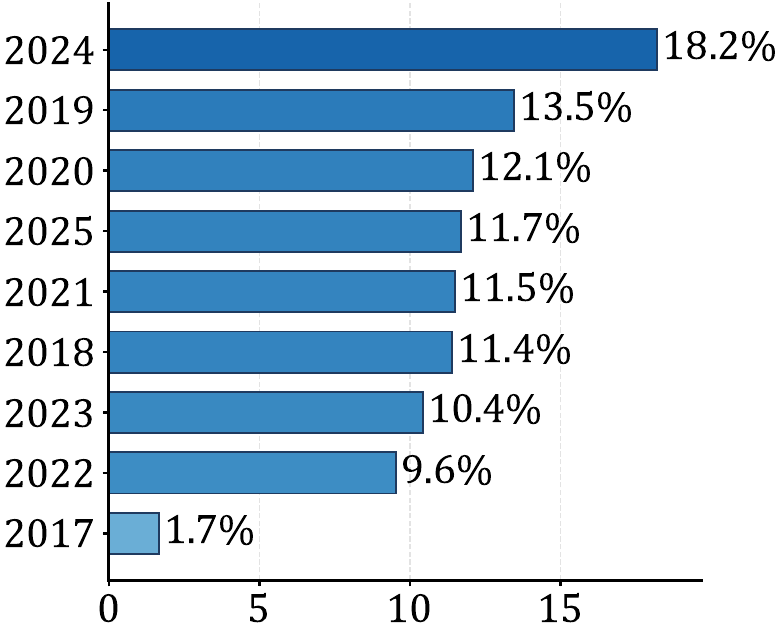}
        \caption{Year}
        \label{fig:year_horizontal_bar}
    \end{subfigure}
    \hfill
    \begin{subfigure}[t]{0.24\textwidth}
        \centering
        \includegraphics[width=\textwidth]{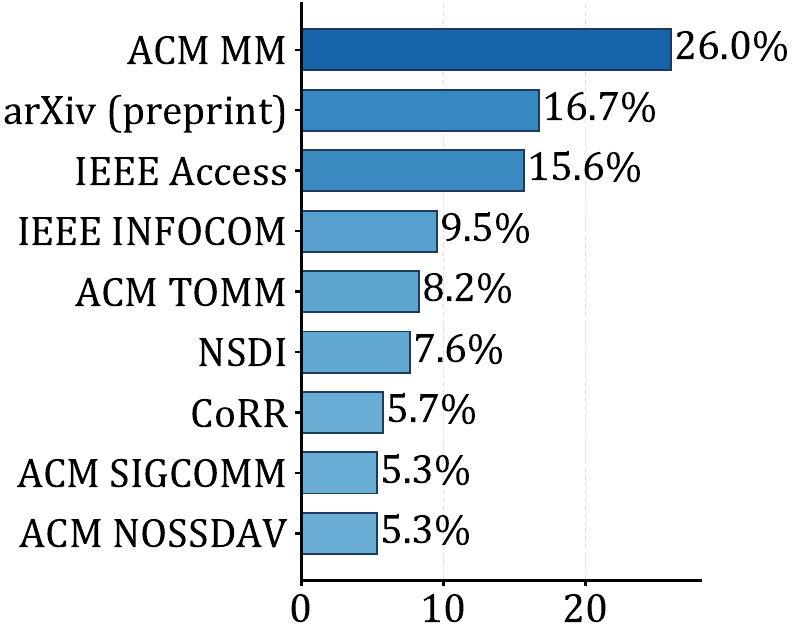}
        \caption{Venue}
        \label{fig:venue_horizontal_bar}
    \end{subfigure}
    \hfill
    \begin{subfigure}[t]{0.24\textwidth}
        \centering
        \includegraphics[width=\textwidth]{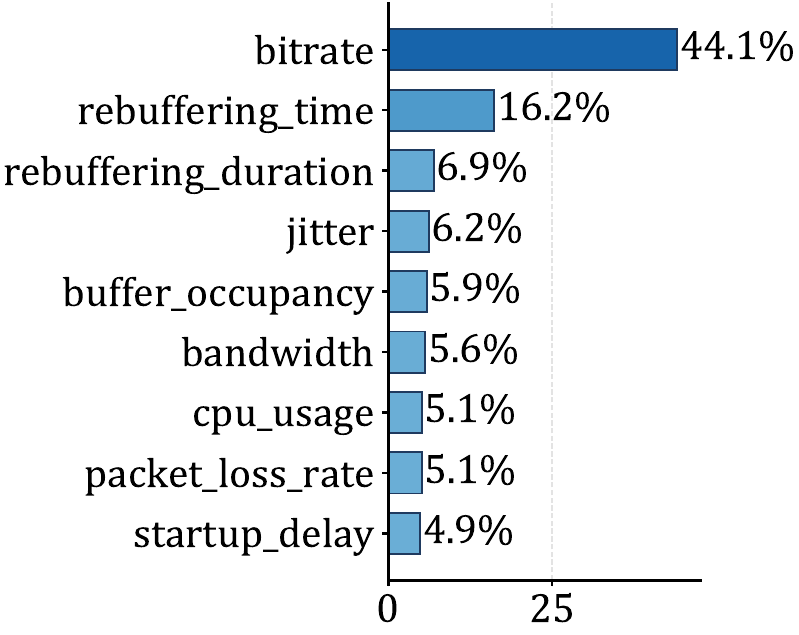}
        \caption{QoS parameters}
        \label{fig:qos_parameter_horizontal_bar}
    \end{subfigure}
    \hfill
    \begin{subfigure}[t]{0.24\textwidth}
        \centering
        \includegraphics[width=\textwidth]{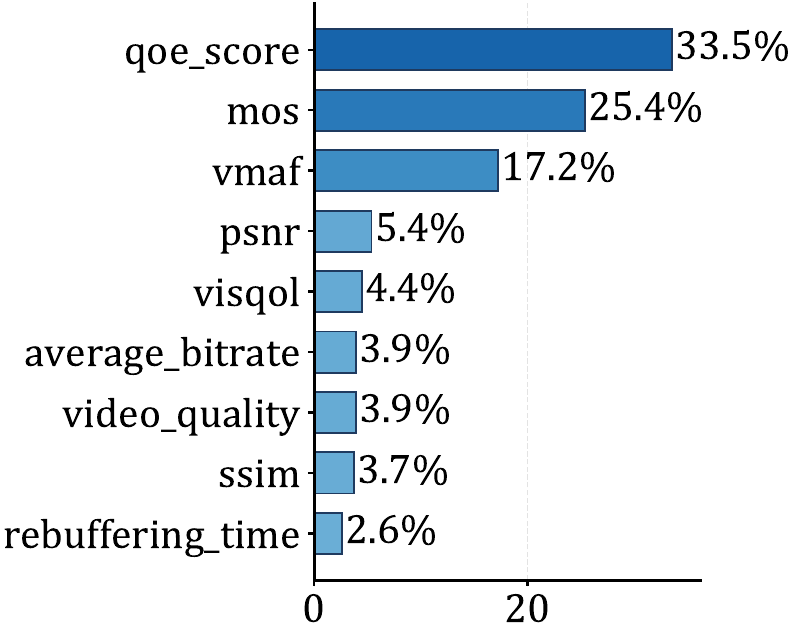}
        \caption{QoE parameters}
        \label{fig:qoe_parameter_horizontal_bar}
    \end{subfigure}

    \caption{Dataset analysis of QoS-QoE Translation. The first row shows key metadata distributions, including protocol, network type, device type, and video type. The second row shows distributions over year, venue, and QoS/QoE parameters.}
    \label{fig:dataset_analysis_all}
\end{figure*}

\textit{QoS-QoE Translation} contains 1026 source-grounded QoS-QoE relationship records extracted from 505 curated papers after extraction and iterative data evaluation. Figure~\ref{fig:data_example} shows an example JSON record, Table~\ref{tab:dataset_field_definitions} summarizes the field definitions, and Figure~\ref{fig:dataset_analysis_all} summarizes the dataset composition in terms of metadata, temporal and venue coverage, and QoS/QoE parameter distributions. Each record contains two main components: \textbf{metadata}, which captures source information and contextual attributes, and \textbf{relationship}, which stores the extracted QoS-QoE relationship. This design preserves contextual information and source-grounded relationships in a unified machine-readable format for downstream analysis, benchmarking, and modeling.

\subsection{Metadata Diversity and Coverage}
Figures~\ref{fig:protocol_horizontal_bar}--\ref{fig:video_type_horizontal_bar} show that the dataset is concentrated in mainstream video streaming settings while still preserving cross-setting diversity. DASH accounts for 59.7\% of the protocol distribution, followed by HTTP at 20.2\%, while WebRTC, QUIC, RTP, and HLS together contribute a non-trivial share of interactive and transport-level settings. For network type, cellular 4G (33.2\%), wired (23.0\%), and Internet-based environments (19.3\%) dominate, indicating that the dataset covers both mobile and fixed-network evaluations. Device type is largely split between desktop (45.2\%) and mobile (41.0\%), suggesting that the dataset mainly reflects common end-user viewing platforms. For video type, 2D video-on-demand is the largest category at 61.3\%, followed by 2D live streaming at 27.8\%, while short video and immersive formats such as 360 and VR video remain represented. Overall, these distributions show that the dataset is anchored in dominant real-world streaming scenarios, but still retains enough diversity to support cross-context analysis.

\subsection{Temporal and Venue Coverage}
Figure~\ref{fig:year_horizontal_bar} shows that the dataset is concentrated in recent years, with 2024 contributing the largest share at 18.2\%, while papers from 2017--2025 remain represented. Figure~\ref{fig:venue_horizontal_bar} further shows that the dataset is collected from a broad set of publication venues. ACM MM contributes the largest share at 26.0\%, followed by arXiv (16.7\%), IEEE Access (15.6\%), IEEE INFOCOM (9.5\%), ACM TOMM (8.2\%), NSDI (7.6\%), and so on. Together, these distributions show that \textit{QoS-QoE Translation} is grounded in both multimedia and networking communities, while maintaining strong coverage of recent research and reducing noise from older system settings.

\subsection{QoS and QoE Parameter Coverage}
Figures~\ref{fig:qos_parameter_horizontal_bar} and~\ref{fig:qoe_parameter_horizontal_bar} show that the dataset covers the most commonly studied QoS and QoE parameters in the literature. On the QoS side, bitrate is the most frequent parameter at 44.1\%, followed by rebuffering time at 16.2\%, while rebuffering duration, jitter, buffer occupancy, bandwidth, CPU usage, packet loss rate, and start up delay each appear in around 5--7\% of records. This pattern shows that the dataset emphasizes system-level factors that are central to adaptive streaming and quality degradation. On the QoE side, QoE score (33.5\%) and MOS (25.4\%) are the most frequent targets, followed by VMAF at 17.2\%, while PSNR, VISQOL, average bitrate, video quality level, SSIM, and rebuffering time appear less frequently. These results indicate that the dataset covers both subjective QoE indicators and objective quality metrics, making it suitable for translation across heterogeneous QoE formulations. Notably, some variables, such as rebuffering time, appear on both the QoS and QoE sides, highlighting that the same concept may be treated differently across studies depending on how the authors define system conditions and user experience outcomes.

\subsection{Dataset Availability and Licensing}
\textit{QoS-QoE Translation} is made publicly available on our project website\footnote{Dataset website: \url{https://yyu6969.github.io/qos-qoe-translation-page/}} to encourage research. The released dataset consists of processed structured JSON records derived from the literature and is distributed under the Creative Commons Attribution 4.0 International (CC BY 4.0) license.
\section{Experiments}
\begin{table*}[t]
    \centering
    \caption{Overall model evaluation results before and after SFT. All values are percentages (\%).}
    \label{tab:main_base_sft_eval}
    \small
    \setlength{\tabcolsep}{5pt}
    \begin{tabular}{lcccccccc}
        \toprule
        \multirow{2}{*}{\textbf{Model}} & \multicolumn{4}{c}{\textbf{Before SFT}} & \multicolumn{4}{c}{\textbf{After SFT}} \\
        \cmidrule(lr){2-5} \cmidrule(lr){6-9}
        & \textbf{MAPE$\downarrow$} & \textbf{Accuracy@$\delta$$\uparrow$} & \textbf{Accuracy$\uparrow$} & \textbf{Macro-F1$\uparrow$}
        & \textbf{MAPE$\downarrow$} & \textbf{Accuracy@$\delta$$\uparrow$} & \textbf{Accuracy$\uparrow$} & \textbf{Macro-F1$\uparrow$} \\
        \midrule
        Qwen3-8B & 20.23 & 64.20 & 67.48 & 55.63 & 11.79 & 78.13 & 80.49 & 72.29 \\
        Qwen3-32B & 16.46 & 67.95 & 68.29 & \underline{56.69} & \underline{9.41} & 80.24 & \underline{84.55} & 70.12 \\
        Qwen3.5-35B-A3B & \underline{14.43} & \textbf{72.94} & \underline{69.92} & \textbf{61.33} & \textbf{8.49} & \textbf{83.41} & \textbf{90.24} & \textbf{84.63} \\
        Llama-3.1-8B-Instruct & 26.72 & 55.90 & 50.41 & 35.62 & 11.34 & 79.49 & 83.74 & 70.79 \\
        Llama-3.3-70B-Instruct & \textbf{13.76} & \underline{72.70} & \textbf{70.73} & 56.29 & 9.49 & \underline{81.90} & \textbf{90.24} & \underline{81.93} \\
        \bottomrule
    \end{tabular}
\end{table*}

\begin{table*}[t]
    \centering
    \caption{Overall model evaluation results after SFT for QoS$\rightarrow$QoE and QoE$\rightarrow$QoS translation. All values are percentages (\%).}
    \label{tab:bidirectional-analysis}
    \small
    \setlength{\tabcolsep}{5pt}
    \begin{tabular}{llcccc}
        \toprule
        \textbf{Task} & \textbf{Model} & \textbf{MAPE (\%)$\downarrow$} & \textbf{Accuracy@$\delta$ (\%)$\uparrow$} & \textbf{Accuracy (\%)$\uparrow$} & \textbf{Macro-F1 (\%)$\uparrow$} \\
        \midrule
        \multirow{6}{*}{QoS $\rightarrow$ QoE}
        & Qwen3-8B & 9.11 & 78.55 & 75.00 & \underline{73.43} \\
        & Qwen3-32B & 7.41 & \underline{82.61} & \underline{77.50} & 62.18 \\
        & Qwen3.5-35B-A3B & \underline{7.05} & \textbf{83.77} & \underline{77.50} & 66.53 \\
        & Llama-3.1-8B-Instruct & 8.79 & 81.74 & 75.00 & 61.31 \\
        & Llama-3.3-70B-Instruct & \textbf{6.88} & \textbf{83.77} & \textbf{85.00} & \textbf{77.78} \\
        \midrule
        \multirow{6}{*}{QoE $\rightarrow$ QoS}
        & Qwen3-8B & 13.47 & 77.67 & 83.13 & 71.43 \\
        & Qwen3-32B & \underline{10.66} & 77.67 & 87.95 & 73.45 \\
        & Qwen3.5-35B-A3B & \textbf{9.40} & \textbf{83.02} & \textbf{96.39} & \textbf{91.81} \\
        & Llama-3.1-8B-Instruct & 12.94 & 77.04 & 87.95 & 73.79 \\
        & Llama-3.3-70B-Instruct & 11.12 & 79.87 & \underline{92.77} & \underline{82.74} \\
        \bottomrule
    \end{tabular}
\end{table*}

The QoS-QoE translation task evaluates whether a model can predict user-experience outcomes from system-level service conditions, or predict system-level service conditions from user-experience observations. The experimental results are summarized in Tables~\ref{tab:main_base_sft_eval} and~\ref{tab:bidirectional-analysis}. The evaluated models are Qwen3-8B, Qwen3-32B, Qwen3.5-35B-A3B, Llama-3.1-8B-Instruct, and Llama-3.3-70B-Instruct~\cite{yang2025qwen3,qwen35_35b_a3b_modelcard,grattafiori2024llama3}. Table~\ref{tab:main_base_sft_eval} reports the overall performance before and after SFT, and Table~\ref{tab:bidirectional-analysis} presents results by translation direction after SFT.

\noindent \textbf{Task Formulation.} We formulate both directions, QoS$\rightarrow$QoE and QoE$\rightarrow$QoS, as structured prediction tasks derived from the source-grounded relationship records in our dataset. In the QoS$\rightarrow$QoE direction, the model predicts user-experience parameters from system-level conditions. In the QoE$\rightarrow$QoS direction, the model predicts system-level conditions from user-experience observations. In both cases, the model receives a structured input instance with contextual information and predicts the queried target field in JSON.

\noindent \textbf{Supervised Fine-tuning.} We use the Tinker framework~\cite{tml2025tinker} for SFT on our QoS-QoE translation tasks. Each example is represented as a multi-turn chat-style JSON instance. The input includes an instruction, task identifier, contextual metadata, scenario description, parameter mapping, source-grounded evidence, history log, and a query, while the target output contains only the predicted JSON field for the queried task.

To construct the SFT corpus, we transform the 1026 source-grounded relationship records into 8107 chat-style instances through holdout-based history reconstruction. For each instance, one target time point from the history log is held out as the query/output pair, and the remaining time points are retained as input context. We use 7205 instances for training and 902 for testing. Unless otherwise noted, all evaluated models use the same split and training configuration. We use the default SFT configuration in Tinker, with a maximum sequence length of 32{,}768 tokens, batch size 128, learning rate $2\times10^{-4}$, a linear learning-rate schedule, and 1 training epoch.

\noindent \textbf{Metrics.} We report four evaluation metrics in our QoS-QoE translation tasks. For continuous value prediction, we use MAPE and Accuracy@$\delta$. MAPE measures the average percentage error between predicted and ground-truth values, where lower is better. Accuracy@$\delta$ measures the fraction of predictions that fall within a predefined tolerance of the ground truth. Because different QoS and QoE parameters have different acceptable error ranges, we use a parameter-specific $\delta$ rather than a single shared threshold. Depending on the parameter, $\delta$ is defined as either an absolute or relative tolerance. The full $\delta$ configuration is provided in the supplementary material. For discrete label prediction, we report Accuracy, which measures exact label matches, and Macro-F1, which averages F1 equally across classes and is less sensitive to class imbalance.

\subsection{Overall Performance}

Table~\ref{tab:main_base_sft_eval} summarizes the overall performance of each model before and after SFT. After fine-tuning, all evaluated models perform reasonably well on both continuous value and discrete label prediction, suggesting that \textit{QoS-QoE Translation} provides a useful supervision signal. Within the same model family, larger models generally outperform smaller ones. For example, Qwen3-32B improves over Qwen3-8B after SFT, reducing MAPE from 11.79\% to 9.41\% and increasing discrete label accuracy from 80.49\% to 84.55\%. Llama-3.3-70B-Instruct also consistently outperforms Llama-3.1-8B-Instruct across all reported post-SFT metrics.

Across model families, Qwen3.5-35B-A3B achieves the strongest overall post-SFT performance. It obtains the best MAPE of 8.49\%, the highest Accuracy@$\delta$ of 83.41\%, ties for the highest discrete label accuracy at 90.24\%, and achieves the best Macro-F1 of 84.63\%. Llama-3.3-70B-Instruct is also highly competitive, reaching 9.49\% MAPE, 81.90\% Accuracy@$\delta$, tying for the best discrete label accuracy at 90.24\%, and obtaining the second-best Macro-F1 of 81.93\%.

Comparing performance before and after SFT, all evaluated models show clear gains on both continuous value and discrete label prediction, indicating that \textit{QoS-QoE Translation} provides effective supervision for bidirectional QoS-QoE translation. For continuous value prediction, Qwen3-8B reduces MAPE from 20.23\% to 11.79\% and improves Accuracy@$\delta$ from 64.20\% to 78.13\%, while Qwen3.5-35B-A3B improves from 14.43\% to 8.49\% in MAPE and from 72.94\% to 83.41\% in Accuracy@$\delta$. Llama-3.1-8B-Instruct shows the largest gain, with MAPE decreasing from 26.72\% to 11.34\% and Accuracy@$\delta$ increasing from 55.90\% to 79.49\%.

The gains are also substantial for discrete label prediction. Qwen3.5-35B-A3B improves from 69.92\% to 90.24\% in accuracy and from 61.33\% to 84.63\% in Macro-F1, while Llama-3.3-70B-Instruct improves from 70.73\% to 90.24\% and from 56.29\% to 81.93\%, respectively. Llama-3.1-8B-Instruct again shows the largest improvement, with accuracy increasing from 50.41\% to 83.74\% and Macro-F1 increasing from 35.62\% to 70.79\%. Overall, these results show that \textit{QoS-QoE Translation} provides an effective supervision signal, leading to strong post-SFT performance and consistent improvements over the corresponding pre-trained baselines.

\subsection{Bidirectional Analysis}

Table~\ref{tab:bidirectional-analysis} reports post-SFT results for the two translation directions separately. Overall, QoS$\rightarrow$QoE appears slightly easier for continuous value prediction, with the best model reaching a MAPE of 6.88\%, compared with 9.40\% for QoE$\rightarrow$QoS. In the QoS$\rightarrow$QoE setting, Llama-3.3-70B-Instruct performs best overall, achieving the lowest MAPE of 6.88\%, tying for the highest Accuracy@$\delta$ of 83.77\%, and obtaining the best discrete label results with 85.00\% accuracy and 77.78\% Macro-F1. In the QoE$\rightarrow$QoS setting, Qwen3.5-35B-A3B performs best overall, with the lowest MAPE of 9.40\%, the highest Accuracy@$\delta$ of 83.02\%, the highest categorical accuracy of 96.39\%, and the highest Macro-F1 of 91.81\%.

The two directions exhibit different patterns. QoE$\rightarrow$QoS is generally more challenging for continuous value prediction, as reflected by higher MAPE values across models, but it yields better discrete label results, especially for Qwen3.5-35B-A3B. One possible explanation is that multiple QoS configurations can correspond to similar QoE outcomes, which makes reverse numeric translation more ambiguous. At the same time, the strong discrete label performance in QoE$\rightarrow$QoS suggests that although precise numeric recovery is harder, coarse-grained reverse translation remains highly learnable.

\section{Related Work}

\subsection{QoS-QoE Modeling and Analysis}

QoS-QoE relationships have been widely studied in multimedia systems \cite{itu_gstr_rq_2023,alreshoodi2013survey}. Prior studies have examined how factors such as bitrate, delay, packet loss, startup latency, stalling, and adaptation behavior affect user-perceived quality across multimedia applications, especially video streaming, using methods including subjective experiments, correlation analysis, analytical modeling, and machine learning-based prediction \cite{barman2019qoe}. Representative studies further showed that video quality impairments and stalling events can strongly affect user engagement and viewer behavior \cite{dobrian2011understanding,krishnan2012video}, and developed predictive models for Internet video QoE using system- and network-level features \cite{balachandran2013developing}. Survey papers have further summarized a broad range of QoS-QoE modeling approaches and challenges in HTTP adaptive streaming \cite{alreshoodi2013survey,barman2019qoe,seufert2015survey}. However, these works mainly focus on understanding or predicting QoE from experimental observations, rather than transforming the published literature itself into a structured and source-grounded dataset for systematic reuse.

\subsection{QoE Datasets and Benchmark Resources}

A separate line of work has produced datasets and benchmark resources for QoE research. Many of these datasets are derived from controlled subjective studies or system-level measurements and provide annotated samples for evaluating QoE prediction models \cite{duanmu2018waterloo,zhu2024taolive,li2023sneset}. They have been valuable for benchmarking and model development, particularly in adaptive video streaming and related applications \cite{duanmu2018waterloo,zhu2024taolive}. However, such resources are typically tied to specific experimental settings, user studies, or measurement campaigns. In contrast, our goal is not to build another experiment-specific QoE benchmark, but to curate a literature-grounded dataset of reported QoS-QoE relationships together with supporting evidence, parameter definitions, and contextual metadata.

\subsection{LLM-based Scientific Document Extraction}

Recent LLMs have shown strong capabilities in information extraction and structured generation from scientific text and documents \cite{shamsabadi2024scientific_ie,dagdelen2024structured_ie}. Multimodal and layout-aware document models extend these abilities to tables, forms, and rich documents, making them promising tools for document understanding and literature mining \cite{wang2024docllm}. Iterative refinement and LLM-based evaluation frameworks have also been explored to improve quality through feedback and aggregation \cite{madaan2023selfrefine,gu2024llmjudge}. Our work builds on these advances, but differs in objective: rather than using LLMs for generic scientific document extraction, we use them to construct a source-grounded QoS-QoE dataset from the literature and combine extraction with iterative multi-reviewer evaluation to improve reliability and traceability.
\section{Discussion}

Although recent large language models have shown strong potential in multimedia applications, important limitations still remain, especially for complex reasoning tasks. In particular, current models are not yet consistently reliable when multi-step inference, temporal understanding, and long-context multimodal reasoning are required~\cite{huang2024survey_eval_mllm,liu2024tempcompass,wu2024longvideobench}. These challenges are especially relevant in multimedia system settings, where useful decisions often depend not only on recognizing visual or audio content, but also on integrating information across modalities and reasoning jointly about system conditions, user experience, and their interactions.
\section{Potential Applications and Impact}

\textit{QoS-QoE Translation} can support several practical applications in video streaming systems and intelligent network management. First, it can serve as a supervision source for training models that predict user-perceived quality from measurable system-level signals such as bitrate, delay, packet loss, or rebuffering. Such models can help service providers estimate QoE in real time without relying only on expensive user studies. Second, the dataset can support reverse prediction from QoE targets to QoS conditions, which is useful for resource planning and adaptive system control, where an operator may want to identify what network or application conditions are needed to achieve a desired level of user experience.

Beyond direct prediction, the dataset also provides a structured knowledge base for retrieval and reasoning. Because each record is source-grounded and linked to equations, tables, or figures in the literature, it can be used in retrieval-augmented systems that answer QoS-QoE questions with evidence and support trustworthy decision making.

More broadly, \textit{QoS-QoE Translation} provides a foundation for LLM-based AI agents for QoS-QoE translation. Such agents could parse user goals, retrieve relevant source-grounded relationships, compare evidence across studies, and generate structured predictions or recommendations for network optimization. They could assist with streaming configuration, quality diagnosis, QoE-aware adaptation, and automatic report generation. Future work will focus on extending the dataset to more diverse and complex scenarios and improving the evaluation framework with more human assessment and finer-grained source analysis.

\section{Conclusion}

In this paper, we present \textit{QoS-QoE Translation}, a source-grounded dataset of QoS-QoE relationships in video streaming. We construct the dataset through a pipeline that combines paper curation, source-grounded relationship extraction, metadata enrichment, and iterative data evaluation, producing machine-readable JSON records with contextual metadata and explicit source traceability. Experiments with supervised fine-tuned large language models show strong performance on both numeric and categorical QoS-QoE translation tasks in both forward and reverse directions, suggesting that \textit{QoS-QoE Translation} is a useful benchmark and training resource for source-grounded QoS-QoE modeling.

\bibliographystyle{ACM-Reference-Format}
\bibliography{sample-base}

\appendix

\end{document}